\documentclass[12pt,onecolumn,prc,amsmath,
floatfix,amssymb]{revtex4}
\usepackage{graphicx}
\begin{document}
\title{On extraction of oscillation parameters}
\author{Jan Sobczyk and Jakub Zmuda}
\affiliation{Institute of Theoretical Physics, University of Wroclaw, \\
plac Maxa Borna 9, 50-204 Wroclaw, Poland}
\date{\today}

\begin{abstract}
We discuss methods to extract neutrino oscillation parameters based on the directly observable quantities, without reconstruction of neutrino energy. The distributions of muon energies and production angles are compared to Monte Carlo predictions made for a set of different neutrino oscillation parameters. The method is applied to T2K neutrino beam and tested for a set of MC data samples in order to evaluate the statistical error. 
\end{abstract}
\maketitle
\section{Motivation}

The traditional method of determination of neutrino oscillation parameters requires reconstruction of the neutrino energy for each measured event. One assumes the target nucleon (mass $M$) to be at rest and bound with mean binding energy $\epsilon_b$. The neutrino energy is then reconstructed from the knowledge of charged lepton kinematics: its scattering angle $\Theta$ and energy $\varepsilon_f$ (or momentum $k_f$):
\begin{equation}
E_{rec}=\frac{\varepsilon_f\left(M-\epsilon_b\right)+ \frac{1}{2}\left(\epsilon_b^2-2M\epsilon_b+m_\mu^2\right)}
{\left(M-\epsilon_b\right)-\varepsilon_f+k_f\cos\Theta}
\end{equation}
The interaction is assumed to be quasi-elastic (QE). The methodology presented in the final K2K experiment analysis is based on likelihood functions, which are dependent on the estimated neutrino energy reconstruction error\cite{Ahn:2006zza}. However at 1 GeV energies the nucleon's Fermi motion and non-QE interactions introduce significant and difficult to control uncertainty \cite{Butkevich:2008ef}.

The T2K experiment is a long baseline accelerator muon disappearance experiment held in Japan. Its main goal is the precise measurement of $\Delta m^2_{23}$ and $\Theta_{23}$ and determination, whether the $\Theta_{13}$ is nonzero. The expected error of the $\Delta m^2_{23}$ is smaller, than 4\% and for the $\Theta_{23}$ is of the order of 1\%. The SuperKamiokande detector would collect about 1600 ${\mathrm{\nu_\mu}}$ cc events/year if there were no oscillation \cite{Loverre:2006rb,Obayashi:2008nj}. The statistics will be large enough to consider different approaches to extract oscillation parameters and in particular to avoid the neutrino energy reconstruction step. Actually, what we observe in SK are mostly muons and (sometimes) pions and all the information about oscillation signal is contained in their distributions. The problem we address is this: which is optimal strategy (e.g. statistical estimator and data binning) to analyse the data? Our considerations and tests were performed for the T2K beam and Super- Kamiokande detection power.
\section{The method proposal}

The method is based on the Monte Carlo prediction of muon distribution in the detector. The events have been produced by the \textit{NuWro} neutrino generator\cite{Juszczak:2005zs}.
The algorithm of oscillation parameter extraction is as follows:
\begin{enumerate}
\item Generate a large number of CC neutrino events with a Monte Carlo generator for the given neutrino beam (a typical size of the sample is 1000000 events).

\item  Create reference oscillation samples for a set of different parameters ($\Delta m^2_{23}$ ,$\sin^2(2\Theta_{23})$) using the muon neutrino survival probability $P(\nu_\mu\rightarrow\nu_\mu )$.

\item Impose detector conditions like {\it muon above Cherenkov threshold}, {\it no observed pions}. Approximate condition for the charged particle being detected is the kinetic energy above the Cherenkov threshold. Neutral pions are assumed to be always visible.

\item Produce independent samples of events of the size corresponding to about
6 years of data gathering. The sampling has been performed by a product of two probabilities: $P(\nu_\mu\rightarrow\nu_\mu )$ (oscillation signal) and 0.01 (normalisation to the right amount of events). This is a way to approximately incorporate the statistical error of the overall normalisation.

Each sample represents a {\it typical} set of events as they are expected to be seen in SK.

\item Compare the experimentally measured muon distribution with the reference samples in order to extract the oscillation parameters. Two strategies are considered:
\begin{itemize}
\item $\mathrm \chi^2$ estimator is used (the considered bins are expected to contain at least 10 or 20 events):
\begin{equation}\label{chi2}
\chi^2=\frac{1}{N_b-2}\sum_{i}\frac{(N_i-n_i)^2}{n_i}
\end{equation}
The index $i$ runs through all bins containing data sets for muon exclusive and muon+pion(s) events. The symbol $N_i$ stands for the number of measured events in each bin whereas $n_i$ is the expected number of events predicted by the Monte Carlo simulation. The normalisation constant $N_b-2$ reflects the fact that our fit model has two unknown parameters: $\sin^2(2\Theta_{23})$ and $\Delta m^2_{23}$. We go through templates produced in step (2) and look for a minimum of the $\chi^2$.

\item Poisson statistical estimator \cite{Ashie:2005ik} is used (the considered bins are expected to contain at least 3 events):
\begin{equation}
F=2\sum_{i}\left[n_i-N_i+N_i\ln{\frac{N_i}{n_i}}\right].
\end{equation}
\end{itemize}
\end{enumerate}

\section{Performance of the method}

The space of oscillation parameters is mapped into a two dimensional lattice on the ($\Delta m^2_{23}$ ,$\sin^2(2\Theta_{23})$) plane. The lattice intervals were $\mathrm{5\times 10^{-5}[eV^2]}$ in $\Delta m^2_{23}$ and $\mathrm{5\times 10^{-3}}$ in $\sin^2(2\Theta_{23})$. Physical expectation constraints have been used on the parameter ranges (e.g. $\sin^2(2\Theta_{23})\leq 1$, $\Delta m^2_{23}\in[21,29]\times 10^{-4}\mathrm{[eV^2]}$).

Tests have been performed for few  values of "true" ($\Delta m^2_{23}$ ,$\sin^2(2\Theta_{23})$). For each one of them 200 MC data samples have been created.

\begin{figure}[ht]\label{fig:result}
\begin{center}
$\begin{array}{cc}
\includegraphics[width=8cm]{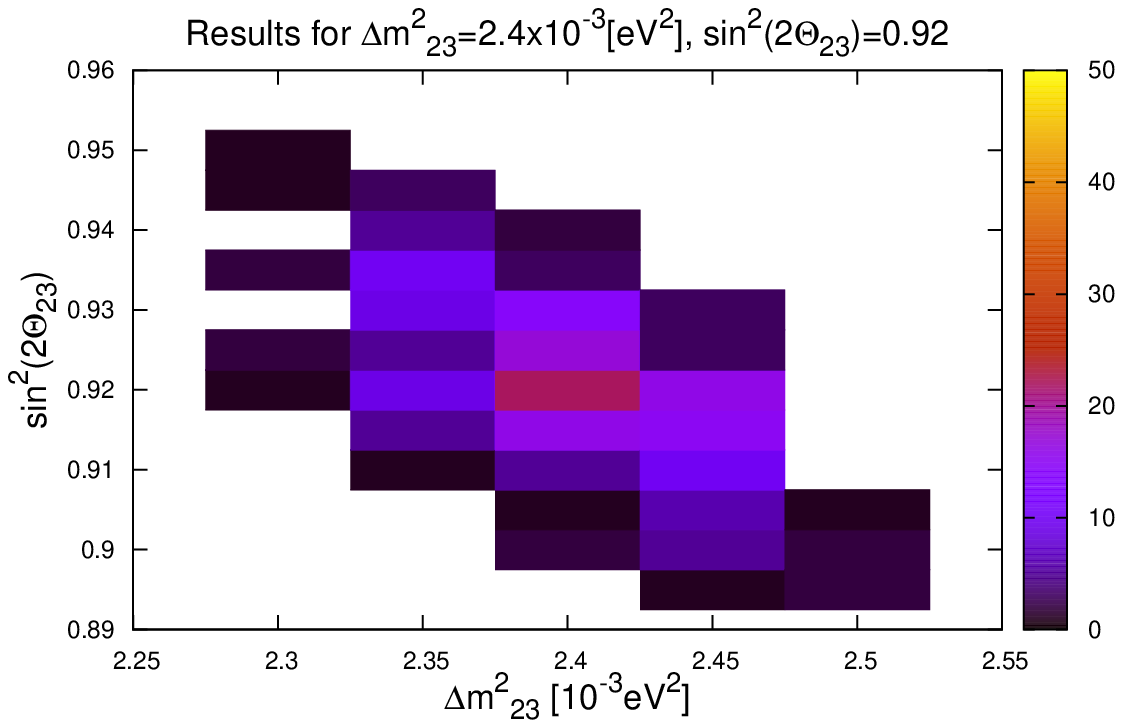} & \includegraphics[width=8cm]{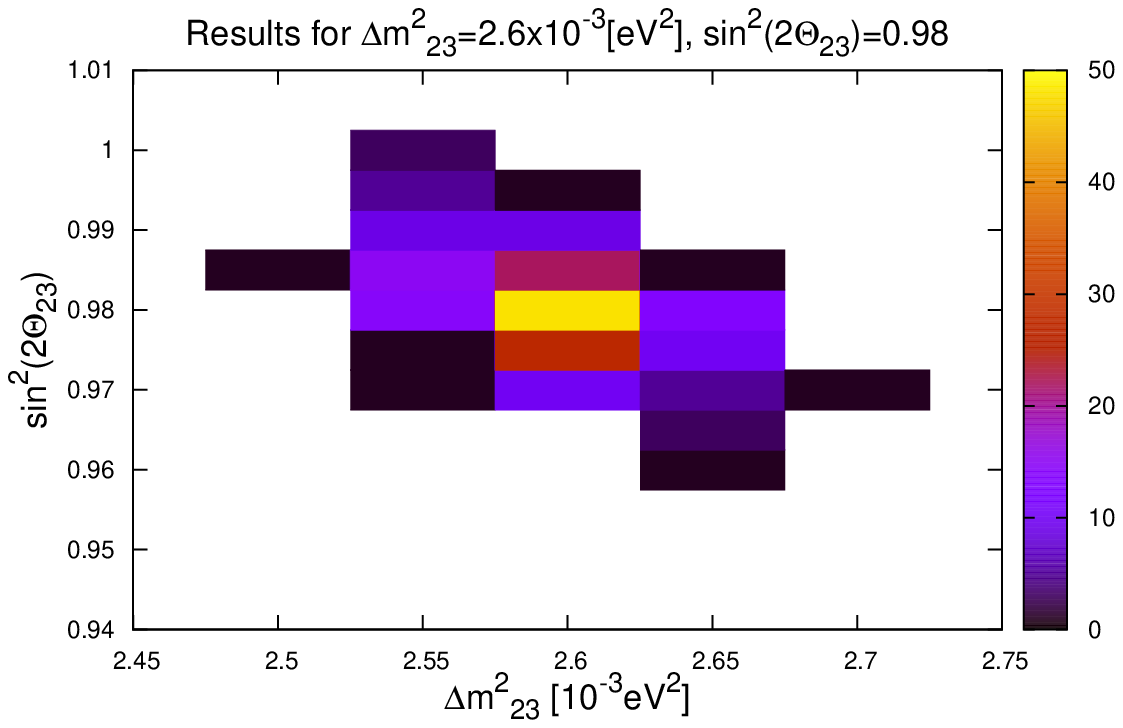}

\end{array}
$
\caption{The figure shows how many of the MC data samples have been identified with each point of the reference lattice.}
\end{center}
\end{figure}

The histograms have been made for several histogram bin sizes and statistical cuts. The range of the bins in muon energy has been fixed to 200-1200 $\mathrm{[MeV]}$.
For the $\chi^2$ statistics 100, 80, 75 and 50 $\mathrm{[MeV]}$ bins have been used. The binning in $\cos(\Theta_{\mu})$ has been varied between 1/4 and 1/20.
The smallest bins were at the verge of Super-K resolution\cite{Ashie:2005ik}.
Alternative binning in $\theta_\mu$ has been also performed. there were nine bins between 0 and 90 degrees and one bin for the backward scattered muons. This binning has been used to extract information in the near 1 [Kt] detector in the Super- Kamiokande experiment\cite{:2008eaa}.
Two statistical cuts were imposed on the bins: at least 10 and at least 20 events. Bins with lower expected statistics had been discarded. Unfortunately, the $\chi^2$ statistics occurred to give unsatisfying results. Too few samples were identified with the right oscillation parameters.
The Poisson method has given better results.The tested bins were 100, 80, 75, 50 $\mathrm{[MeV]}$ and 1/5, 1/10, 1/15, 1/20 $\mathrm{\left[\cos(\Theta_{\mu})\right]}$. The best result has been found for 50 $\mathrm{[MeV]}$ x 1/5 $\mathrm{\left[\cos(\Theta_{\mu})\right]}$ binning.
The figure \ref{fig:result} shows the distribution of results on the reference lattice for ($\Delta m^2_{23}=24\times 10^{-4}[eV^2]$ ,$\sin^2(2\Theta_{23})=0.92$) and ($\Delta m^2_{23}=26\times 10^{-4}[eV^2]$ ,$\sin^2(2\Theta_{23})=0.96$). Each bin on the plot gives the number of data samples identified with one point of the lattice.

\section{Conclusions}
The method gives quite good concentration of the results around the expected value. Judging from the plots, without inclusion of the systematic errors, results have 1 $\sigma$ areas of about $\pm 0.5\times 10^{-4}\mathrm{[eV^2]}$ in $\Delta m^2_{23}$ and $\pm 0.02$ in $\sin^2(2\Theta_{23})$. The error in $\Delta m^2_{23}$ is close to the expected uncertainties of T2K\cite{Loverre:2006rb,Obayashi:2008nj}, but the uncertainty of $\sin^2(2\Theta_{23})$ is larger. These uncertainties depend strongly on the area, in which the true values of oscillation parameters are located. For small values of both $\sin^2(2\Theta_{23})$ and $\Delta m^2_{23}$ the concentration of the results around true value is rather poor. It seems, that this method is better at extracting the squared mass difference then the mixing angle. This happens probably due to the error of overall normalisation. The position of the oscillation probability maximum depends on $\Delta m^2$, whereas its depth depends on the $\sin^2(2\Theta)$. Thus the total number of recorded events should depend more on the $\sin^2(2\Theta)$, then $\Delta m^2$.

More effort must be made to find optimal method of binning. Perhaps using non regular bin shapes will improve the method's performance. In general a compromise must be found: to have many bins in the region sensitive to oscillation signal but also to keep high statistics.

We are aware that the systematic errors will add more uncertainty but its evaluation is a separate complicated problem.

\section*{Acknowledgements}

The authors were supported by the grant 35/N-T2K/2007/0 (the project number DWM/57/T2K/2007). We thank Federico Sanchez for a discussion.

\end{document}